# Transport properties of very overdoped nonsuperconducting Bi$_2$Sr$_2$CuO$_{6+\delta}$ thin films

H. Raffy,* Z.Z. Li, P. Auban-Senzier

Laboratoire de Physique des Solides, CNRS UMR 8502, Bâtiment 510, Université Paris-Saclay, 91405 Orsay, France.

The transport properties, resistance, Hall effect, and low T magnetoresistance for very oxygen overdoped nonsuperconducting Bi$_2$Sr$_2$CuO$_{6+d}$ (Bi2201) thin films are reported. From 20 to 300K, the temperature dependence of the resistance is well described by a law of the form $a+bT^{4/3}$, theoretically predicted to occur in the presence of ferromagnetic fluctuations. In addition, this prediction is reinforced by the analysis of the transverse and the longitudinal low T magnetoresistance. Interestingly, the presence of a weak disorder causing low T electronic localization allows us to evidence very short diffusion lengths, as observed in other systems with ferromagnetic fluctuations.

## I. INTRODUCTION.

After a long period during which the studies on cuprates essentially focused on the underdoped and optimally doped regions of their phase diagram (PD), recent studies are considering the overdoped side and particularly the end of the overdoped region. In the underdoped side, it is well known that superconductivity appears by doping an antiferromagnetic insulating phase. Increasing the doping, the critical temperature, T$_c$, undergoes a maximum at optimal doping and decreases in the over doped side to disappear for very over doped (VOD) cuprates, yielding a pure metallic state. This behavior raises a question regarding the reason for which superconductivity disappears with increasing doping or inversely why superconductivity appears with decreasing it. Trying to answer this question constitutes another way for unveiling the mystery of the pairing in the cuprates. This point has been recently considered theoretically [1, 2]. Maier *et al.* [1] attribute the right end of the T$_c$ dome to the combined effect of the decreasing *d* wave pairing strength, closely coupled to changes in the magnetic fluctuations, and to weak impurity scattering. Although they think reasonable to associate the drop of superconductivity to a weakening of pairing interaction, Li *et al.* [2] consider that the reduction of the superfluid density may be the cause of the superconductor-to-metal transition. For these authors, the key ingredient is a combination of *d*-wave pairing and disorder, leading to a granular superconductor. Experimentally, the loss of superconductivity by doping the cuprates can only be observed in one-layer cuprates, i.e., in La$_{2-x}$Sr$_x$CuO$_4$ [3], Tl$_2$Ba$_2$CuO$_2$ [4], and Bi$_2$Sr$_2$CuO$_6$ (Bi2201)] [5] considered in this study. In Bi2201, nonsuperconducting metallic samples can be obtained either by cationic substitution Bi/Pb, giving different samples for different cationic substitution, or by oxygen doping of a given thin film [5].The latter technique allows one to observe the evolution of the transport properties on one and the same sample [6]. The VOD part of the PD is somewhat simpler than the less doped one: the (PG) disappears for a given doping $p^*$ (considered as a quantum critical point [7], $p^*$ being smaller than the doping where superconductivity dies out; scanning tunneling microscopy (STM) experiments have shown [8] that charge order and nematicity, reported when superconductivity exists, vanish at high doping when superconductivity has disappeared. At the same time, nanoscale patches of static charge order with $\sqrt{2} \times \sqrt{2}$ periodicity are observed. Angle-resolved photoemission spectroscopy (ARPES) experiments on Pb doped-Bi2201 single crystals [9,10.] have shown that the evolution of the Fermi surface (FS) topology with doping exhibits a Lifshitz transition from hole-like FS to *e*-like FS with increasing doping. However, while Ding *et al.* [9] conclude that this transition coincides with the loss of superconductivity, Valla *et al.* [10] report that it occurs at higher doping than that where superconductivity disappears. Furthermore, as discussed already in Ref. [11], the obtained metallic state of VOD does not show the canonical $T^2$ Fermi liquid behavior. The linear temperature dependence of the resistivity, existing in the whole temperature range at optimal doping, still exists at low temperature with increasing doping within a smaller temperature range. One qualifies this behavior as strange metal.

Another characteristic of the cuprates is the presence of antiferromagnetism which survives in a large part of the PD, from the doped Mott insulator to the overdoped state where the PG disappears [12]. Unexpectedly, it has been predicted theoretically that ferromagnetic fluctuations appear in the nonsuperconducting overdoped (OD) state [13]. Several experiments were conducted to look for ferromagnetic fluctuations in different overdoped (OD) non superconducting cuprates: in OD La$_{2-x}$Sr$_x$CuO$_4$ single crystals by the technique of muon spin relaxation (μSR) [14.]; in a OD Bi2201 single crystal by magnetization and muon experiments [15] and also in *e*-doped La$_{1-x}$Ce$_x$CuO$_4$ thin films by magnetoresistance, magnetic measurements and Kerr effect [16]. These different studies concluded either to the presence of a dilute concentration of static magnetic moments [14] or to the presence of two-dimensional (2D) ferromagnetic fluctuations [15], or to the emergence of itinerant ferromagnetism [16].

---

*Corresponding author: helene.raffy@universite-paris-saclay.fr

In order to give a complementary characterization of the VOD region of the PD of cuprates, we have studied the transport properties (Hall Effect and low $T$ magnetoresistance) of Bi2201 thin films at high oxygen doping state for which superconductivity is suppressed. The present study complements the study previously published on Bi(La)2201 with $T_{cmax}$ equal to 30K at optimal doping and $T_c$= 15K for these maximally oxygen doped Bi(La)2201 samples [6] due to the presence of La. For VOD thin films presented here, in the absence of superconductivity, a slight increase of the resistance $R(T)$ and of the Hall constant $R_H(T)$ appears at the lowest temperatures suggesting a possible presence of disorder. It was shown that the low $T$ magnetoresistance is very anisotropic which will be ascribed to $e$-$e$ interaction effects and weak electronic localization. Although we could not observe ferromagnetism directly in our transport measurements, magnetoresistance is a probe of the hole dynamics which can disclose a behavior involving spins.

## II. EXPERIMENTAL TECHNIQUES

Pure Bi2201 thin films, with a thickness ranging 100-350 nm, were epitaxially grown by a sputtering technique on $c$-axis oriented SrTiO$_3$ single crystal substrates. X-ray diffraction (XRD) studies allowed us to show that these films are single phase and $c$-axis oriented. Our technique of film synthesis as well as the film characterization were described in our earlier references [5,6,17]. We have shown in Ref. [17] typical XRD characterization results (XRD $\theta$-$2\theta$ scan and XRD $\phi$ scan in Figs. 4 and 5 of Ref.[17] respectively) as well as a cross section of such Bi2201 films grown on SrTiO$_3$ substrates (in Fig. 1 of Ref. [17]).

In order to reach by oxygenation the pure metallic phase required for this study, we chose to synthetize thin films with a maximum critical temperature $T_{cmax}$ around 10 K. $T_{cmax}$ of Bi2201 thin films is essentially depending on the cationic composition, which was found to be Bi$_{2.1}$Sr$_{1.85}$CuO$_y$ for the thin films studied here (the same $T_c$ value was reported for thin films prepared by molecular beam epitaxy [18] with comparable cationic composition, superconducting Bi2201 thin films being slightly Sr deficient).

The films have been patterned mechanically into a Hall bar, parallel to the $a$-crystallographic axis, equipped with six gold contacts for $R(T)$ and Hall measurements. They were overdoped by an oxygen treatment as previously reported [5]. The maximum oxygen doping is obtained by annealing in oxygen plasma.

Transport properties, resistance, Hall effect and magnetoresistance, $R(T)$, $R_H(T)$, $R(H,T)$ have been measured in a Quantum Design physical property measurement system (PPMS) (2K$\leq$T<400K; 0$\leq$H$\leq$9T). One sample was measured down to 40 mK using an adiabatic demagnetization refrigerator adapted to the PPMS setup. The temperature dependence of the Hall coefficient $R_H(T)$ has been deduced from the measurements of the magnetoresistance performed successively at 9 and -9T with: $R_H(T)$= ($r$(T,9T)-$r$(T,-9T))/2$H$, and $r$ the resistance measured on Hall contacts.

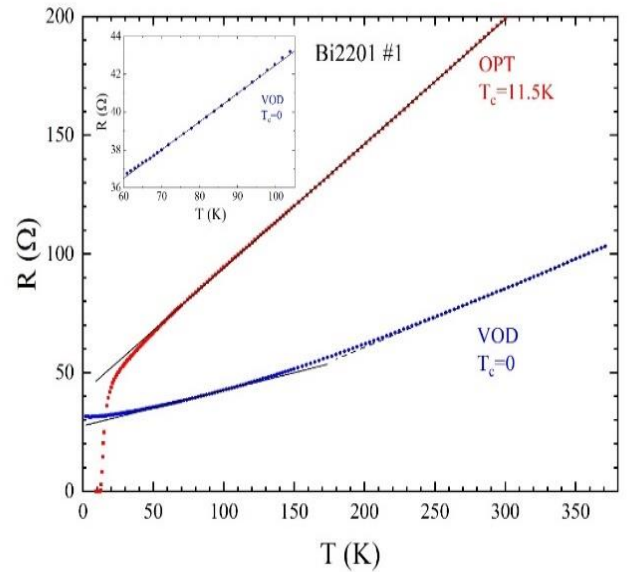

FIG. 1. Temperature dependence of the resistance of thin film Bi2201 no.1 measured at optimal (OPT) doping (red symbols) and after oxygenation in a VOD state (black symbols). The lines are fits to the linear parts of the $R(T)$ curves: at optimal doping from 50 to 300K (continuous line); in the VOD state, at low $T$ from 65K to 100K (continuous line and inset) and at high $T$ for 250 to 370K (dashed line).

## III. RESULTS AND DISCUSSION

### A. Resistance vs T

As shown in our previous studies [6], the increase of oxygen content in a Bi2201 thin film appears clearly in the shape of both $R(T)$ and (see next paragraph) $R_H(T)$. Figure 1 displays two typical $R(T)$ curves obtained for Bi2201 no.1 thin film at optimal doping and after plasma oxidation. For the optimally doped state the resistance $R(T)$ is linear in temperature down to the superconducting fluctuations before the superconducting transition ($T_c$=11.5K). The $R(T)$ curve of the VOD state exhibits lower resistance values at given $T$, a well-marked increase of the curvature, and no superconducting transition.

Looking for the strange and bad metal behavior of cuprate resistance occurring at low $T$ and high $T$ respectively, one can extract from this latter curve two linear parts (see lines in Fig. 1). The former at low $T$ occurs for 66K$\leq$T$\leq$100K (see inset) with a slope of $R(T)$ equal to 0.149 $\Omega$/K and the latter at high $T$ is observed for 250K$\leq$T$\leq$370K with a slope equal to 0.247 $\Omega$/K corresponding respectively to slopes of the resistivity equal to 0.805 and 1.334 $\mu\Omega$.cm/K. From the low $T$ slope and half the distance of CuO$_2$ planes $c$/2=12.33$\pm$0.02 Å, we obtain a value of the slope of the resistivity per Cu plane: A$_\square$=6.5$\pm$0.3, comparable to the Planckian limit for Bi2201 [7]: 8$\pm$2.

Actually, the shape of $R(T)$ shown in Fig. 2 can be well described by a law of the type: $a+bT^n$. One finds $n$=1.40 for 20K$\leq$T$\leq$300 K and $n$=1.37 for 20 K$\leq$T$\leq$370 K (see the red line fit in Fig. 2). One may wonder if a linear variation of $R(T)$ replaces the $T^{4/3}$ law at higher $T$. We did not check it above 370K as heating could induce a loss of oxygen and a change of the doping state of the thin film.

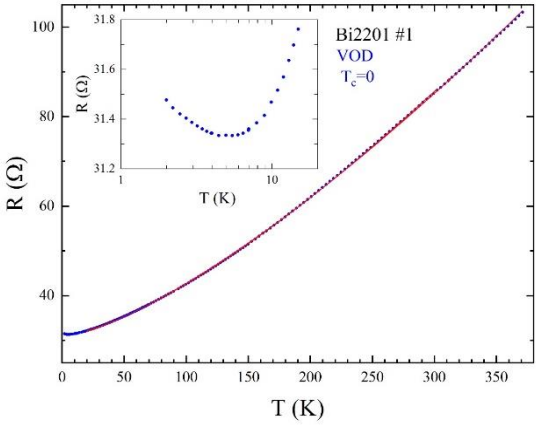
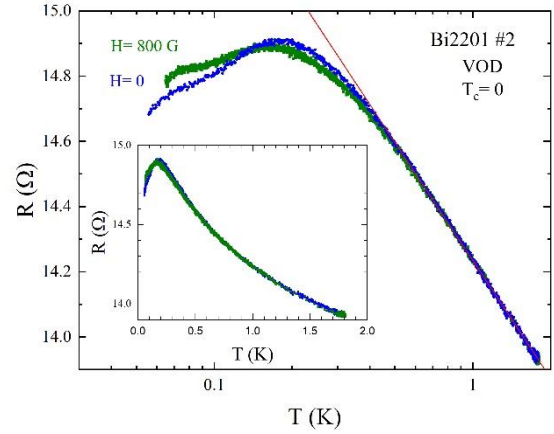

FIG. 2. Detailed behavior of the $R(T)$ curve given in Fig.1 for thin film Bi2201 no 1 in the VOD state. The $R(T)$ curve is fitted by a law of the form: $R(T) = a+bT^n$ above 20 K, $n$= 1.4 up to 300 K, and $n$=1.37 up to 370 K (red lines). The inset visualizes the logarithmic increase of $R(T)$ below 10 K

FIG. 3. $R(T)$ for 40 mK<$T$<2 K for VOD Bi2201 no. 2 thin film in a semilog plot (blue curve) showing the logarithmic variation of $R(T)$ (red line). The green curve is a plot of $R(T, 800G)$ showing the influence of a perpendicular magnetic field $H$=800 G. Inset: the same plots in a linear $T$ scale display the maximum of $R(T)$ occurring for $T$=180 mK, revealing the presence of some superconducting islands.

For the different VOD Bi2201 thin films of the present study, one obtains 1.33≤ $n$ ≤1.4 for 20 K ≤$T$≤ 300 K. It was proposed theoretically that an exponent 4/3 could be present in a 2D conductor presenting ferromagnetic fluctuations [19]. Our result reinforces recent μSR experiments, magnetization and $R(T)$ measurements which have been made on VOD Bi(Pb) 2201 single crystals, which indicate the presence of ferromagnetic fluctuations [15] in nonsuperconducting Bi2201. The fact that $R(T)$ in thin films and single crystal behave the same confirms the quality of the thin films studied. Since the beginning of our studies on BiSrCuO thin films, we have systematically considered the evolution of the exponent $n$ with doping and reported the value $n$=1.3 for non superconducting VOD Bi2201 thin films [20] as well as for VOD La doped thin films still superconducting ($T_{cmin}$ = 17K)[6]. The present paper is an attempt to give an explanation to these values as well as those obtained in our early studies.

As shown in the inset of Fig.2, an increase of $R(T)$ is observed at low $T$, below a minimum located between 6 and 10 K for the different Bi2201 films studied with comparable $T_{cmax}$ values [21]. We attribute this increase to the presence of a weak disorder. We will see in the following that the study of the magnetoresistance in presence of disorder allows us to obtain interesting information on the scattering process present in these films.

In order to observe this resistance increase and to check the absence of superconducting transition at lower temperature, the resistance of Bi2201 no. 2 thin film was measured down to 40 mK with the special equipment mentioned in Sec. II. Below 2K, $R(T)$ increases logarithmically before reaching a broad maximum at 180 mK followed by a slight decrease down to 40 mK (Fig. 3). Applying a magnetic field of H = 800G (the maximum field allowed by our cooling system) $R(T, 800 G)$ is first slightly lower than $R(T, 0)$ in the increasing part of $R$(T), then below the maximum, $R(T, 800 G)$ becomes higher than $R(T, 0)$ in the decreasing part of $R(T)$.

The latter may be an indication partial destruction of superconducting fluctuations associated to remaining superconducting islands, responsible for the observed maximum of $R(T)$ The former may be attributed to a partial destruction of the weak electronic localization (see Sec.III C.2).

### B. Hall effect

It is well known that the Hall effect is not constant vs $T$ in cuprates and that it does not bring the same information as in classical metal [22]. The temperature dependence of the Hall coefficient, $R_H(T)$, has been measured in a magnetic field of ±9T at optimal doping and at VOD oxygen doping successively on the same film. $R_H(T)$ increases from 300 K to a maximum occurring below T=100 K. The comparison of both curves $R_H(T)$ obtained for the same film as in Fig.1, plotted in Fig. 4, shows sthat the carrier concentration is increased by oxygen doping. As, at optimal doping, superconductivity is still present under $H$=9 T, $R_H(T)$ decreases to zero when approaching $T_c$. In the case of a VOD nonsuperconducting state, $R_H(T)$ is still temperature dependent, with a maximum around 100 K as in the optimal doping situation and as it was already reported for all doping states [23,24]. Remarkably enough, a low $T$ increase of the Hall constant is observed for all our nonsuperconducting VOD thin films. It can be seen in Fig. 4 and even more clearly in Fig. 5 for another VOD Bi2201 thin film. It occurs in the $T$ region where $R(T)$ presents an increase attributed to localization and $e$-$e$ interaction effects. While the low T $R(T)$ increase results both from $e$ localization and $e$-$e$ interaction, the low $T$ increase of the Hall constant is only ascribed to $e$-$e$ interaction effect as predicted by the interaction model [25], and the localization model [26] predicts that $R_H$ is not modified by the $e$ localization. A low $T$ increase of the Hall constant has been reported for underdoped LaSrCuO single crystals[1] showing also an increase of R(T) at low T.

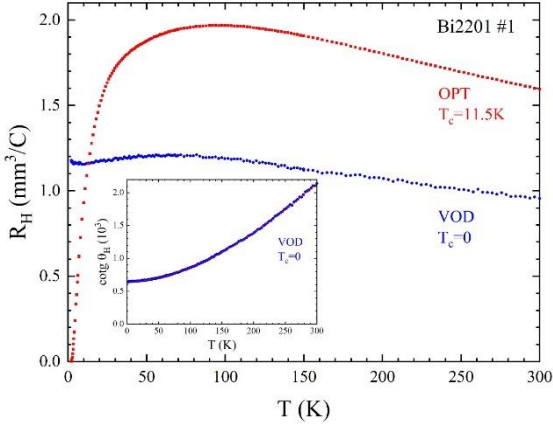

FIG. 4. Temperature dependence of the Hall constant measured under $H = 9$ T of thin film Bi2201 no 1, in red at optimal (OPT) doping and in blue in the VOD state. At $T = 300$ K, the Hall numbers are respectively: $n_H$(OPT) = (3.9±0.2)x$10^{21}$/cm$^3$ and $n_H$(VOD) = (6.5±0.3)x$10^{21}$/cm$^3$. Note the slight increase of $R_H$ below 10 K in the VOD state. Inset: $T$ dependence of cotg $\theta_H = \rho/H.R_H$ (blue points). The red line is the fit of the data with a law of the type: $A+BT^m$ ($m = 1.73$).

The inset displays the $T$ dependence of the Hall cotangent. It associates the resistivity $\rho$ and the Hall constant $R_H$ as defined by

$$\cot \theta_H = \rho / HR_H$$

For $2\,K \leq T \leq 300\,K$, a law of the form: $A+BT^m$ can fit the data with $m=1.73$, still far from 2 [22].

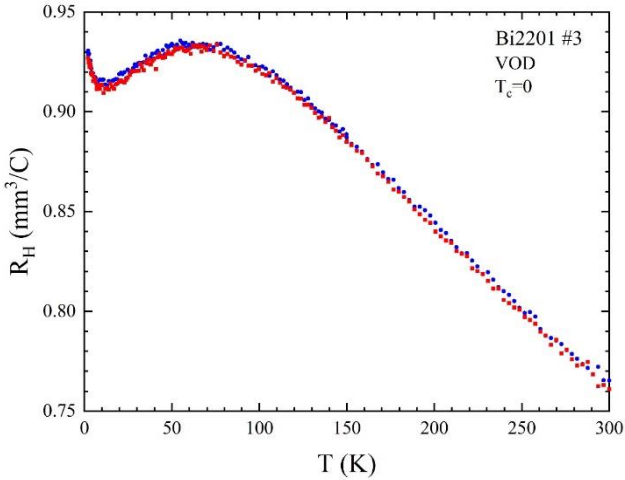

FIG. 5. Temperature dependence of the Hall constant $R_H$(T) of VOD thin film Bi2201 no. 3 showing clearly the low $T$ increase of $R_H$(T). The measurements made with both pairs of Hall contacts (blue and red points) are given.

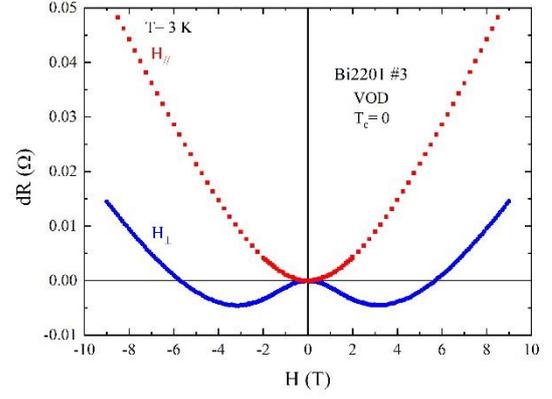

FIG. 6. Magnetoresistance measured in perpendicular and parallel magnetic field at $T = 3$ K for VOD Bi2201 no. 3 thin film.

As discussed in Ref. [22.] in the conventional transport theory, the Hall number in the low $T$ and high field limit is simply related to the number of charge carriers, $n_H$. It was also shown [22] that for Bi2201 the field dependence of $R_H$ in the normal state is weak (Fig. S3 of Ref. [22]). From the value of $R_H$ measured at our lowest temperature $T = 2$ K under a field $H = 9$ T, the hole concentration per volume unit $n_H = 1/eR_H$ can be derived although the small increase of $R_H$ due to $e$-$e$ interaction reduces slightly the value $n_H$. For the sample of Fig. 5, $n_H$ is found equal to (6.72±0.05) x $10^{21}$/cm$^3$, while $n_H$ = (6.85±0.05) x $10^{21}$/cm$^3$ at T= 10 K where $R_H$(T) is minimum. The number of holes per Cu, $n_{Cu}$, is obtained using the volume V of the unit cell of Bi2201 and the number of Cu atoms in the unit cell [28]. It is found equal to 1.27±0.1, giving a value of the doping per Cu: p = $n_{Cu}$-1= 0.27±0.1.

### C. Magnetoresistance R(H, T)

The magnetoresistance (MR) was measured from 2 to 25 K in a perpendicular magnetic field $H_\perp$ (transverse MR) and in a magnetic field parallel to the current $H_{//}$ (longitudinal MR) for $0 \leq H \leq 9T$ (and also in one case in a transverse field with an angle close to 45° in a plane perpendicular to the film containing the current $I$.

Figure 6 presents a typical result obtained at 3 K in perpendicular and parallel magnetic fields for VOD Bi2201 no. 3 thin film. The longitudinal MR is positive and increases steadily with $H$. On the contrary, the transverse MR is negative at low magnetic field, becoming positive at high magnetic field. It is also very different from the MR measured when the sample is less doped and still superconducting. In that case the MR is positive, dominated by the suppression of the superconducting fluctuations. These two situations are described in detail in the following paragraphs.

#### 1. Longitudinal MR

The magnetic field being parallel to the current, one expects the longitudinal magnetoresistance to arise from the coupling of the spins to the magnetic field $H$. In two

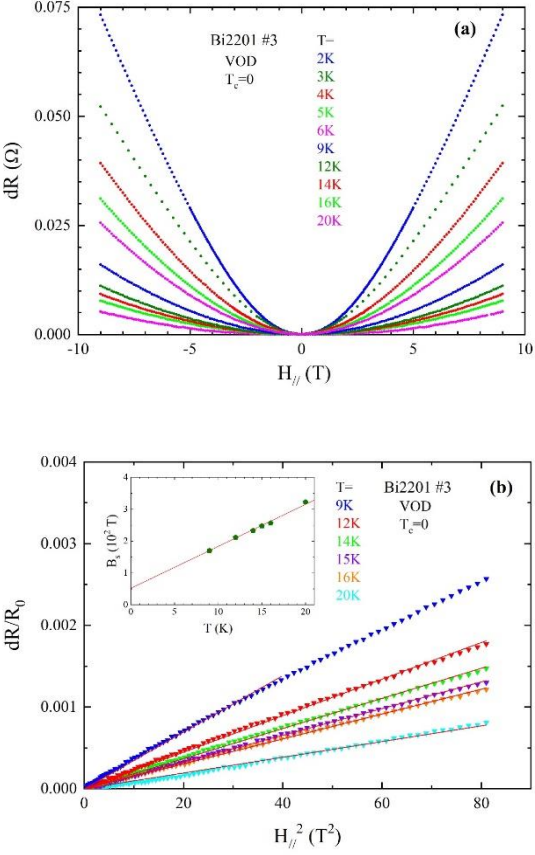

FIG. 7.(a) Magnetoresistance of VOD thin film Bi2201 no 3: $dR=R(H, T) - R(0, T)$ measured in a field parallel to the current at various $T$ (2-20 K).

(b) Plot of $dR/R(0)$ vs $H^2_{//}$ for $T \geq 9$ K, where $dR$ is equal to $R(H)-R(0)$, and linear fits giving the values of the characteristic field $B_s$ defined as $dR/R(0)=(B/B_s)^2$. The inset shows the $T$ dependence of $B_s$ and the linear fit.

dimensions, while the orbital contribution is sensitive only to the magnetic field component normal to the plane, the spin splitting term should be isotropic [29] (no spin orbit).

Figure 7(a) displays $R(H_{//})$ data obtained on a VOD Bi2201 thin film measured at various $T$. In a magnetic field parallel to the current, $R(H_{//})$ is always positive in the domain of temperature study (2 K ≤ $T$ <25 K). The field dependence of $R(H_{//})$ is quadratic at low field: $H \leq 4T$ for the lowest $T$ values and up to 9 T for $T \geq 10$ K [Fig. 7(b)] where the data are fitted by a parabolic law of the form

$$dR/R(0) = (H/B_s)^2$$

where $R(0)$ is the value of the resistance in zero field with $dR = R(H)-R(0)$ and $B_S$ a characteristic field related to $e$-$e$ interaction.

If the magnetoresistance was due to the Zeeman effect, as it can be thought with a field parallel to the current, one would expect a parabolic law of the form

$$dR/R(0 \approx (H/B_z)^2$$

with $B_z = k_B T / g\mu_B = 0.744$ T

The values of $B_z$ extracted from the parabolic fits are plotted in the inset of Fig. 7(b). Although $B_z (T)$ is linear, the linear fit does not go through the origin, and its slope (13.2T) is about 20 times higher than that of the Zeeman field $B_z$ (0.744 T/K). Besides, we do not observe any saturation of the MR with increasing field at the lowest temperatures for which the MR stays linear at the highest field values.

In contrast to the transverse orbital MR (see next paragraph), the longitudinal MR ($H//I$) could not be analyzed with existing theories. At this point it is interesting to compare these results to experimental ones obtained in previous studies reporting the MR of $Pd_{1-y}Ni_yH_x$ thin films near the paramagnetic-ferromagnetic crossover [30]. The $Pd_{1-y}Ni_yH_x$ alloys are ferromagnetic for y≥ 2.5 % Ni and x = 0 and become less magnetic and even superconducting when hydrogenated to saturation ($x$ ~ 1). In these films the approach of the ferromagnetic state was monitored by increasing y and decreasing x from the saturated state. It was shown, for paramagnetic states already close to be ferromagnetic ($y$ = 16 at %, $x$ = 1), that the MR is positive and parabolic in longitudinal magnetic field and also in perpendicular magnetic field with smaller MR values, the presence of localization being no longer seen (state very close to a ferromagnetism). Increasing the magnetism by decreasing $x$ ($x$<1) for a given $y$, the MR becomes isotropic. In our Bi2201 thin films the localization is still present in the transverse MR, but the transverse MR tends to become positive at high magnetic field. The longitudinal magnetoresistance follows a parabolic variation as described above and the field $B_s(T)$ presents a $T$ linear variation as observed for a characteristic field $H_N$ for spin splitting (Fig.5 of Ref.[30]). The similar behavior of the longitudinal MR of Bi2201 and $Pd_{1-y}Ni_yH_x$ thin films suggests that the longitudinal magnetoresistance may result from the presence of ferromagnetic fluctuations in VOD Bi2201 thin films.

At low T, the parabolic variation of $R(H,T)$ is followed by a linear variation with increasing fields. It would be interesting to measure the MR at higher magnetic fields to see if one still observes a linear variation of $R(H)$ without saturation as reported in perpendicular and parallel magnetic field in Bi2201 single crystals and discussed in [31].

### 2. *Perpendicular field MR: What is the origin of the negative MR observed at low field?*

The low T upturn of $R(T)$ and the negative part of the perpendicular field MR recall the behavior observed in conventional metal thin films in the weak localization regime [32, 33]. For instance, in Cu thin films [34], there is a low $T$ increase of $R(T)$ and a negative transverse MR, attributed at low field to the localization of the carriers in presence of disorder. The application of a perpendicular magnetic field destroys the interference effect of the conduction electrons leading to a negative MR in the absence of spin-orbit effect

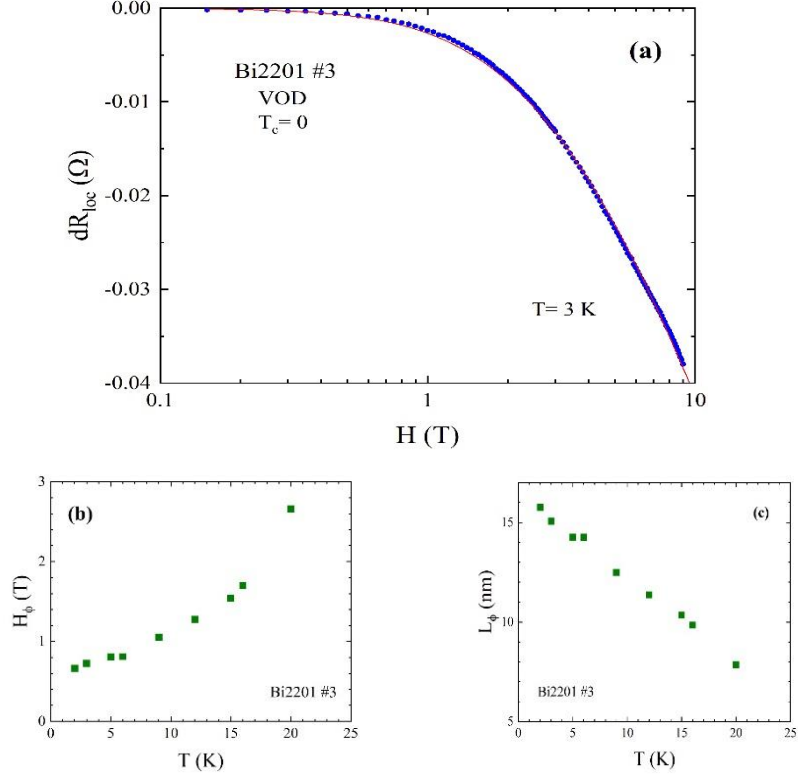

FIG. 8 (a) Negative contribution or orbital part of the magnetoresistance of VOD Bi2201 no. 3 thin film at $T$ = 3 K in perpendicular magnetic field fitted by the expression of the MR due to weak localization (see text). The parameters obtained from the fit are: $H_\Phi$ (3 K) = (0.7261±0.0035) T and $L_\Phi$(3 K) = (15±0.5) nm. (b) Temperature dependence of the characteristic dephasing field $H_\Phi$ and (c) of the dephasing length $L_\Phi$.

(MR becoming positive in the latter case). Such an interpretation of the MR in terms of weak localization was made early in [35] to analyze the low $T$ MR of non superconducting Bi2201 single crystals [nonequivalent to our VOD thin films with different $R(T)$ and $R_H(T)$ dependence) presenting a low $T$ increase of $R(T)$. We also assume that the longitudinal MR only involves $e$-$e$ interactions, which are isotropic. So the orbital part of the transverse MR, $dR_{loc}$, is obtained by subtracting the longitudinal MR $dR_{//}$ from the transverse one $dR_\perp$, so that

$$dR_{loc} = dR_\perp - dR_{//} \text{ where } dR = R(H) - R(0).$$

We have analyzed the orbital part of the MR measured at low $T$ using the expression [36]

$$dR_{loc}(H)/RR_\square = -(\alpha e^2/2\pi^2 \hbar) Y[H/H_\Phi] \quad (1)$$

where $Y(x) = \ln x + \Psi(1/2 + 1/x)$,

$H_\Phi = \hbar/4eD\tau_\Phi$ is the dephasing magnetic field

$L_\Phi = \sqrt{D\tau_\Phi}$ is the dephasing length

with $D$ the diffusion constant and $\tau_\Phi$ the dephasing time

$R_\square$ is the resistance per square

$\Psi(x)$ is the digamma function showing a quadratic behavior at low $x$ and a $\ln x$ behavior at high $x$ values; $\alpha$ is a constant of the order of unity in the absence of spin orbit effect.

We have fitted our data with expression (1) between 2 and 20 K. Good fits are obtained for the samples studied. Figure 8 shows a typical result of the fit for the orbital component of the MR at $T$=3K shown in Fig. 6.

Our analysis of the orbital MR is supported by measurements we made in inclined magnetic field with an angle $\theta_D$ ($\theta_D = 40°$) with the film plane for which the orbital part of the MR obtained in the same way by subtracting the $e$-$e$ contribution and plotted vs the perpendicular component of the magnetic field $H\sin\theta_D$ follows the same law as the MR measured in perpendicular magnetic field: at a given $T$, the curves representing $dR_{loc}(H \sin\theta_D)$ vs $H \sin\theta_D$ and $dR_{loc}(H)$ vs $H$ are well superimposed for $H \leq H\sin\theta_D$.

Figures 8(b) and 8(c) display respectively the temperature dependence of the dephasing magnetic field $H_\Phi(T)$ and the dephasing length $L_\Phi$ for the same sample. $H_\Phi(T)$ is increasing with $T$, almost linearly at low $T$, and its value is about ten times higher than that observed in ordinary metal films [32].

The values of the dephasing lengths in our work are very comparable to those reported by Jing et al.[35] for their non superconducting Bi2201 single crystal. On the other hand, the transport mean free path $l$, for instance at 5K, is found equal to 4 nm (with $k_F l$ = 30 and $k_F$ = 7.22.10$^7$ cm$^{-1}$ the 2D Fermi vector, deduced from resistivity and Hall constant values).

As indicated in the Introduction, the presence of ferromagnetic fluctuations in VOD cuprates has been predicted theoretically [13. 19] and reported experimentally in VOD Bi2201 single crystals [15]. In that case, it was deduced from magnetization measurements and μSR experiments. For thin films, it is difficult to deduce from such experiments valuable information concerning the presence of magnetic fluctuations (too small material quantity, presence of magnetic impurities in the substrate). However, it can be thought that the large values of $H_\Phi(T)$ obtained for our VOD non superconducting thin films are possibly related to the presence of ferromagnetic fluctuations.

As in the case of the longitudinal MR, we come to this conclusion by considering the experimental results of the preceding work concerning the transverse MR in Pd, PdH [37] and $Pd_{1-y}Ni_yH_x$ (y = 6 at%) [38] thin films. Approaching in a controlled manner a ferromagnetic state by decreasing x from 1, the analysis of 2D $e$ localization present in the transverse MR of these films allowed the authors to identify the signature of magnetic fluctuations in these systems. It was shown that the dephasing magnetic fields were larger in pure Pd than in PdH. This effect was related to the tendency for Pd to be closer to a magnetic instability (see Fig.1 in Ref. [37]) and attributed to the scattering from extended magnetic fluctuations. The study of weak $e$ localization in 2D $Pd_{1-y}Ni_yH_x$ thin film (y = 6 at%, Ref. [38])) showed that a strong increase of $H_\Phi$ in comparison with $H_\Phi$ in PdH thin films is observed for $x = 1$ while a decrease of $x$ leads a further increase of $H_\Phi$ (see Fig 5 of Ref. [38]). These effects were related to the growth of ferromagnetic fluctuations.

Relying on the results of these experimental studies, we propose to attribute the large values of $H_\Phi(T)$ in VOD Bi2201 to the presence of ferromagnetic fluctuations.

## IV. SUMMARY

The transport properties of Bi2201 thin films in a VOD nonsuperconducting state obtained by plasma oxygenation have been compared to the transport properties of the *same* samples in an optimally doped state. The temperature dependence of the resistance of VOD thin films follows a law of the form $a+bT^{4/3}$ in the range 20-300 K, which was predicted to reveal ferromagnetic fluctuations [13,15]. At low $T$, the metallic state of these nonsuperconducting cuprates do not present the canonical Fermi liquid behavior. Contrariwise, a low $T$ logarithmic increase of $R(T)$ is observed. Such an increase, also seen under high magnetic field in less doped or optimally doped states of the same samples, is attributed to the presence of a weak disorder in Bi2201 thin films. The Hall constant in the VOD state, still $T$ dependent, with smaller values than for the optimal state, is consistent with high doping. An interesting observation is the low $T$ increase of $R_H(T)$ in the same $T$ region where $R(T)$ increases. We attribute the $T$ increase of $R_H(T)$ to $e$-$e$ interaction only, and the resistivity increase to $e$-$e$ interaction and weak electronic localization.

Interestingly, the study of the combined effects of magnetic field and disorder at low $T$ allowed us to probe the importance of excitations existing in this temperature and doping range. The magnetoresistance is anisotropic. The longitudinal magnetoresistance is positive whatever $H$ ($0 \leq H \leq 9T$) and $T$ ($2K \leq T \leq 25K$). It is attributed to $e$-$e$ interaction. The analysis of its parabolic law may suggest the presence of ferromagnetic fluctuations. The transverse magnetoresistance of VOD Bi2201, drastically different from the one in the superconducting states, is negative at low field becoming positive at high field. It is supposed to result from a superposition of the isotropic $e$-$e$ contribution and orbital 2D weak localization. The analysis of the orbital magnetoresistance in terms of 2D weak localization allows us to determine the dephasing fields $H_\Phi(T)$. The high values of these characteristic fields compared to those reported for classical metallic thin films reveal a large scattering effect which may indicate the presence of ferromagnetic fluctuations. Our conclusions are supported by a comparison of our results with those of previous studies of $Pd_{1-y}Ni_yH_x$ [30, 38] thin films near the paramagnetic-ferromagnetic cross-over.

More studies would be necessary to understand why superconductivity appears with decreasing doping and how ferromagnetic fluctuations give way to anti-ferromagnetic ones at lower doping when the pseudogap is opened. Spin excitations appears to be present in the different parts of the PD of cuprates and may play a role in their superconductivity [39].

## ACKNOWLEDGMENTS

H. R. thanks M. Gabay and C. Pasquier at LPS, Université Paris-Saclay, and P. Monceau at CNRS, Institut Néel, Grenoble, for stimulating and very helpful discussions.